\documentstyle[12pt]{article}
\newlength{\extraspace}
\newlength{\extraspaces}
\newcommand{\be}{\begin{equation}
\addtolength{\abovedisplayskip}{\extraspaces}
\addtolength{\belowdisplayskip}{\extraspaces}
\addtolength{\abovedisplayshortskip}{\extraspace}
\addtolength{\belowdisplayshortskip}{\extraspace}}
\newcommand{\ee}{\end{equation}}

\newcommand{\ba}{\begin{eqnarray}
\addtolength{\abovedisplayskip}{\extraspaces}
\addtolength{\belowdisplayskip}{\extraspaces}
\addtolength{\abovedisplayshortskip}{\extraspace}
\addtolength{\belowdisplayshortskip}{\extraspace}}
\newcommand{\ea}{\end{eqnarray}}

\newcommand{\nonu}{\nonumber \\[.5mm]}
\newcommand{\A}{&\!\!\!}

\newcommand{\newsection}[1]{
\vspace{7mm} \pagebreak[3] \addtocounter{section}{1}
\setcounter{subsection}{0} \setcounter{footnote}{0}
\begin{center}
{\large {\bf \thesection. #1}}
\end{center}
\nopagebreak
\medskip
\nopagebreak \hspace{3mm}}

\begin{document}
%%%%%%%%%%%%%%%%%%%%%%%%%%%%%%%%%%%%%%%%%%%%%%%%%%%%%%%%%%%%%%%%%%%%%%%%%%

\begin{center}
{{\bf Energy and angular momentum  of general 4-dimensional
stationary axi-symmetric spacetime in teleparallel geometry}}
\end{center}
\centerline{ Gamal G.L. Nashed}

\bigskip

\centerline{{\it Mathematics Department, Faculty of Science, Ain
Shams University, Cairo, Egypt }}

\bigskip
 \centerline{ e-mail:nashed@asunet.shams.edu.eg}

\hspace{2cm}
\\
\\
\\
\\

We derive an exact  general axi-symmetric solution  of the coupled
gravitational and electromagnetic fields  in the tetrad theory of
gravitation. The solution  is characterized by four parameters $M$
(mass), $Q$ (charge), $a$ (rotation) and $L$ (NUT). We then,
calculate the total exterior energy using the energy-momentum
complex given by M\o ller in the framework of Weitzenb$\ddot{o}$ck
geometry. We show that the energy contained in a sphere is shared
by its interior as well as exterior. We also calculate the
components of the spatial momentum to evaluate the angular
momentum distribution. We show that the only non-vanishing
components of the angular momentum is in the Z direction.

\begin{center}
\newsection{\bf Introduction}
\end{center}

 A general stationary axi- symmetric object, with gravitomagnetic monopole
and dipole moments associated with nonzero values of the NUT and
Kerr parameters $L$ and $a$  respectively, is described by
Kerr-NUT (Newman-Unti-Tamburino) spacetime \cite{NUT}
 which is  a useful model for exploring gravitomagnetism \cite{BC}. The Kerr-NUT
spacetime and its spacial  cases are all belong to larger class of
stationary axi-symmetric type D. Carter \cite{Cb} has found vacuum
solutions of  Einstein equations  for which the Hamilton-Jacobi
equation of the geodesic is separable. The stability of the
general axi-symmetric spacetime is probed  by studying their
perturbation by fields of various spin \cite{BD,BDC}.

Einstein's general relativity (GR) is a very successful theory in
describing long distance phenomena. However, it encounters from
serious difficulties on short distances. The main problem appears
in all attempts is the quantization \cite{Sc,Yi}. Also, the
Lagrangian structure of GR differs from the ordinary microscopic
gauge theories. In particular, a covariant conserved
energy-momentum tensor for the gravitational field can not be
constructed in the framework of GR. Consequently, the study of
alternative models of gravity is justified from the physical as
well as from the mathematical point of view. Even in the case when
GR is a unique true theory of gravity, consideration of close
alternative models can shed light on the properties of GR itself.

It is well known that in GR an energy-momentum tensor  fails to
satisfy some certain conditions \cite{Yi1}. This is usually
related to the equivalence principle which implies that the
gravitational field can not be detected at a point as a covariant
object. This  can  be viewed as a purely differential-geometric
fact. Since the components of the metric tensor are managed by a
system of second order partial differential equations therefore,
the energy momentum quantity has to be a local tensor constructed
out from the metric components and their first order derivatives.
The corresponding theorem of (pseudo) Riemannian geometry states
that every expression of such type is trivial. It is natural to
expect that this objection for the existence of a gravitational
energy-momentum tensor is directly related to the geometric
properties of the (pseudo) Riemannian manifold. This objection can
be lifted in alternative model, even connected with the geometry
of the manifold.

In recent time teleparallel structures on spacetime have evoked a
considerable interest for various reasons. They were considered as
an essential part of generalized non-Riemannian theories such as
the Poincar$\acute{e}$ gauge theory ( \cite{HNH} $\sim$ \cite{BN}
and references therein) or metric-affine gravity \cite{HMM}.
Physics relevant to geometry may be related to the teleparallel
description of gravity \cite{HS1,NH}. Teleparallel approach is
used for positive-gravitational-energy proof \cite{Me1}. The
relation between spinor Lagrangian and teleparallel theory is
established \cite{TN}.

M\o ller has shown that the problem of the energy-momentum complex
has no solution in the framework of gravitational field theories
based on Riemannian spacetime \cite{Mo}. In a series of papers,
\cite{Mo,Mo3,Mo4} he was able to obtain
 a general expression for a satisfactory energy-momentum complex in the absolute
 parallelism space. The Lagrangian formulation
of the theory was given by Pellegrini and Plebanski \cite{PP}.
Quite independently Hayashi and Nakano \cite{HN} formulated the
tetrad theory of gravitation as a gauge theory of the spacetime
translation group. In these attempts, the admissible Lagrangians
were limited by the assumption that the field equations has the
Schwarzschild solution. M\o ller \cite{Mo1} abandoning this
assumption and look for a wider class of Lagrangians by
constructing a new field theory. His aim was to get a theory free
from singularity while retaining the main merits of GR as far as
possible. Meyer \cite{Me} showed that M\o ller's theory is a
special case of the Poincar$\acute{e}$ gauge theory \cite{HS,HNV}.
S$\acute{a}$ez \cite{Se} generalized M\o ller theory into a scalar
tetradic theory of gravitation.

The tetrad theory  of gravitation  based on the geometry of
absolute parallelism \cite{PP}$\sim$\cite{AGP} can be considered
as the closest alternative to GR and it has a number of attractive
features both from the geometrical and physical viewpoints.
Absolute parallelism is naturally formulated by gauging spacetime
translations. Translations are closely related to the group of
general coordinate transformations which underlies GR. Therefore,
the energy-momentum tensor represents the matter source in the
field equation for the gravitational field is just like in GR.

The tetrad formulation of gravitation was considered by M\o ller
in connection with attempts to define the energy of gravitational
field \cite{Mo1,Mo2}. For a satisfactory description of the total
energy of an isolated system it is necessary that the
energy-density of the gravitational field is given in terms of
first- and/or second-order derivatives of the gravitational field
variables. It is well-known that there exists no covariant,
nontrivial expression constructed out of the metric tensor.
However, covariant expressions that contain a quadratic form of
first-order derivatives of the tetrad field are feasible. Thus it
is legitimate to conjecture that the difficulties regarding the
problem of defining the gravitational energy-momentum are related
to the geometrical description of the gravitational field rather
than are an intrinsic drawback of the theory \cite{Mj,MDTC}.

The main aim of the present paper is to derive a general
axi-symmetric solution in the tetrad theory of gravitation for the
coupled gravitational and electromagnetic fields.  Using this
solution we then, calculate the energy and angular momentum using
the e{\it nergy-momentum complex given by M\o ller \cite{Mo1} and
Mikhail et.al \cite{MWHL}}. In \S 2 we derive the field equations
of the coupled gravitational and electromagnetic fields. In \S 3
we obtain a new exact analytic axi-symmetric  solution
characterized by four parameters in the tetrad theory of
gravitation. In \S 4 we calculate the energy and angular momentum
distribution of this solution using M\o ller's energy-momentum
complex \cite{Mo1}. The final section is devoted to discussion and
conclusion.

\newsection{Tetrad theory of gravitation and electromagnetism}

In the Weitzenb{\rm $\ddot{o}$}ck geometry the fundamental field
variables describing gravity are a quadruplet of parallel vector
fields \cite{HS} ${e_i}^\mu$\footnote{Latin indices
$(i,j,k,\cdots)$ designate the vector number, which runs from
$(0)$ to
 $(3)$, while Greek indices $(\mu,\nu,\rho, \cdots)$ designate the world-vector components
running from 0 to 3. The spatial part of Latin indices is denoted
by $(a,b,c,\cdots)$, while that of Greek indices by $(\alpha,
\beta,\gamma,\cdots)$.}, which we call the tetrad field in this
paper, characterized by \be D_{\nu} {e_i}^\mu=\partial_{\nu}
{e_i}^\mu+{\Gamma^\mu}_{\lambda \nu} {e_i}^\lambda=0\; , \ee where
${\Gamma^\mu}_{\lambda \nu}$ defines the nonsymmetric affine
connection coefficients. The metric tensor $g_{\mu \nu}$ is given
by \[g_{\mu \nu}= \eta_{i j} {e^i}_\mu {e^j}_\nu,\] with the
Minkowski metric $\eta_{i j}=\textrm {diag}(+1\; ,-1\; ,-1\; ,-1)$
\footnote{ Latin indices are rasing and lowering with the aid of
$\eta_{i j}$ and $\eta^{ i j}$.}.  The curvature tensor defined by
${\Gamma^\lambda}_{\mu \nu}$, given by Eq. (1) i.e.
${R^\sigma}_{\mu \nu \lambda}(\Gamma)$, is identically vanishing
\cite{HS}.

 The gravitational Lagrangian $L_G$ is an invariant constructed
 from $g_{\mu \nu}$ and the contorsion tensor
  $\gamma_{\mu \nu \rho}$ given by \be \gamma_{\mu \nu \rho} =
{e^i}_{\mu}e_{i \nu; \ \rho}  \;, \ee where the semicolon denotes
covariant differentiation with respect to Christoffel symbols. The
most general gravitational Lagrangian density invariant under
parity operation is given by the form \cite{HN,HS, MWHL}
 \be
{\cal
 L}(e)_G  =  b L_G(e) = e \left( \alpha_1 \Phi^\mu \Phi_\mu
 + \alpha_2 \gamma^{\mu \nu
\rho} \gamma_{\mu \nu \rho}+ \alpha_3 \gamma^{\mu \nu \rho}
\gamma_{\rho \nu \mu} \right) \ee
 and $\Phi_\mu$ being the basic vector field defined by
 \be \Phi_\mu = {\gamma^\rho}_{\mu \rho},\ee  where $e=det({e^a}_\mu)$.
  Here $\alpha_1\; ,
\alpha_2\; ,$ and $\alpha_3$ are constants determined
 such that the theory coincides with general relativity in the weak
 fields \cite{HN,Mo1}:
\be
 \alpha_1=-{1 \over \kappa}\; , \qquad \alpha_2={\lambda \over
\kappa}\; , \qquad \alpha_3={1 \over \kappa}(1-\lambda)\; , \ee
 where
$\kappa$ is the Einstein constant and  $\lambda$ is a free
dimensionless parameter\footnote{Throughout this paper we use the
relativistic units$\;$ , $c=G=1$ and
 $\kappa=8\pi$.}. The vanishing of this parameter makes the theory
 coincides with GR formulated in teleparallel geometry.

The electromagnetic Lagrangian  density ${\it L_{e.m.}}$ is
\cite{KT}
 \be {\it L_{e.m.}}=-\displaystyle{1 \over 4} g^{\mu \rho}
g^{\nu \sigma} F_{\mu \nu} F_{\rho \sigma}\; , \ with \ F_{\mu
\nu} \ being \ given \ by\ \footnote{Heaviside-Lorentz
rationalized units will be used.} F_{\mu \nu}=
\partial_\mu A_\nu-\partial_\nu A_\mu, \ee where $A_\mu$ is the vector
potential. The variation of the gravitational Lagrangian given by
Eq. (3) with respect to the tetrad field  gives one equation of
motion. M\o ller assumed that the energy-momentum tensor of matter
fields, i.e., $T_{\mu \nu}$ is symmetric. Therefore M\o ller
divided the resulting field equation into symmetric part and skew
symmetric part \cite{Mo1}.

The gravitational and electromagnetic field equations for the
system described by ${\it L_G}+{\it L_{e.m.}}$ are the following:

 \be G_{\mu \nu} +H_{\mu \nu} =
-{\kappa} T_{\mu \nu}\; , \ee \be K_{\mu \nu}=0\; , \ee \be
\partial_\nu \left( \sqrt{-g} F^{\mu \nu} \right)=0 \ee
with $G_{\mu \nu}$ being the Einstein tensor of GR. Here
 $H_{\mu \nu}$ and $K_{\mu \nu}$ are defined by \be H_{\mu \nu}
\stackrel {\rm def.}{=} \lambda \left[ \gamma_{\rho \sigma \mu}
{\gamma^{\rho \sigma}}_\nu+\gamma_{\rho \sigma \mu}
{\gamma_\nu}^{\rho \sigma}+\gamma_{\rho \sigma \nu}
{\gamma_\mu}^{\rho \sigma}+g_{\mu \nu} \left( \gamma_{\rho \sigma
\lambda} \gamma^{\lambda \sigma \rho}-{1 \over 2} \gamma_{\rho
\sigma \lambda} \gamma^{\rho \sigma \lambda} \right) \right]\; ,
 \ee
and \be K_{\mu \nu} \stackrel {\rm def.}{=} \lambda \left[
\Phi_{\mu\; ,\nu}-\Phi_{\nu\; ,\mu} -\Phi_\rho
\left({\gamma^\rho}_{\mu \nu}-{\gamma^\rho}_{\nu \mu} \right)+
{{\gamma_{\mu \nu}}^{\rho}}_{;\rho} \right]\; , \ee and they are
symmetric and antisymmetric tensors, respectively. The
energy-momentum tensor $T^{\mu \nu}$ is given by \be T_{\mu
\nu}=-g^{\rho \sigma}F_{\mu \rho}F_{\nu \sigma}+\displaystyle{1
\over 4} g_{\mu \nu} F^{\rho \sigma} F_{\rho \sigma}. \ee

It can be shown \cite{HS} that the tensors, $H_{\mu \nu}$ and
 $K_{\mu \nu}$, consist of only those terms which are linear or quadratic
in the axial-vector part of the torsion tensor, $a_\mu$, defined
by \be a_\mu \stackrel{\rm def.}{=} {1 \over 3} \epsilon_{\mu \nu
\rho \sigma} \gamma^{\nu \rho \sigma}, \ee where $\epsilon_{\mu
\nu \rho \sigma}$ is defined by \be \epsilon_{\mu \nu \rho \sigma}
\stackrel{\rm def.}{=} \sqrt{-g} \delta_{\mu \nu \rho \sigma}, \ee
with $\delta_{\mu \nu \rho \sigma}$ being completely antisymmetric
and normalized as $\delta_{0123}=-1$. Therefore, both $H_{\mu
\nu}$ and $F_{\mu \nu}$ vanish if the $a_\mu$ is vanishing. In
other words, when the $a_\mu$ is found to vanish from the
antisymmetric part of the field equations, (8), the symmetric part
(7) coincides with the Einstein equation formulated in the
Weitzenb$\ddot{o}$ck spacetime.
\newsection{ Exact Analytic Solutions}

In this section we will seek for a solution satisfying the
following conditions: The parallel vector fields having the form
\be {e_\mu}^k={\delta_\mu}^k+Ml_\mu l^k+\displaystyle{Q^2-L^2
\over 2}m_\mu m^k.\ee Here $M$, $Q^2$ and $L^2$ are free
parameters, $l_\mu$ and $m_\mu$ are quantities satisfying the
conditions \be \eta^{\mu \nu}l_\mu l_\nu=0, \qquad \qquad
\eta^{\mu \nu}m_\mu m_\nu=0, \qquad \qquad  \eta^{\mu \nu}l_\mu
m_\nu=0, \ee and $l^k$ and $m^k$ are defined by \be l^k \stackrel
{\rm def.}{=} {\delta^k}_\mu \eta^{\mu \nu} l_\nu \; , \qquad
\qquad m^k \stackrel {\rm def.}{=}{\delta^k}_\mu \eta^{\mu \nu}
m_\nu.\ee Applying (15) to the field equations (7)$\sim$(9) we
obtain the values of $l_\mu$ and  $m_\mu$ in the form \be
l_0=\sqrt{\Xi}\; , \qquad l_\alpha=\displaystyle {2\sqrt{\Xi}
\over \Upsilon+r^2+h^2}\left[\Omega x_\alpha+\displaystyle
{a^2x_3{\delta^3}_\alpha \over \Omega}-\epsilon_{\alpha \beta 3} a
x^\beta \right ], \qquad m_\mu=\displaystyle {l_\mu \over
\sqrt{\Omega}}\; ,\ee where \be \Xi=\displaystyle {\Omega \over
\Upsilon}\; , \qquad \Upsilon=\sqrt{(r^2-a^2)^2+4a^2z^2}\; ,
\qquad \Omega=\displaystyle {\sqrt{r^2-a^2+\Upsilon \over 2}}\; ,
\ee with $r=\sqrt{x^2+y^2+z^2}$ and $\epsilon_{\alpha \beta
\gamma}$ are the three dimensional totally antisymmetric tensor
with $\epsilon_{1 2 3}=1$ and $a$ is the angular momentum per unit
mass. The form of  the  electromagnetic potential $A_\mu$ has the
expression \be A_\mu=\frac{-q}{4\pi} \ \sqrt{\Xi} \ l_\mu, \ \ and
\ \  the \ \ Maxwell \ \ field \ \  F_{\mu \nu}=\frac{q}{4\pi}
\left[\left(\sqrt{\Xi} \ l_\nu \right)_{,\  \mu} -\left(\sqrt{\Xi}
l_\mu \right)_{, \ \nu}\right],\ee where $\Xi$ and $l_\mu$ are
defined in Eqs. (18) and (19).

For the solution given by Eqs. (18) and (20), the axial vector
part $a_\mu$ of the torsion tensor vanishes, \be a_\mu=0,\ee and
the metric is identical to  Kerr-NUT metric in GR.

Writing explicitly the tetrad (15) using (18)  and (19) we
obtain\footnote{We will denote the symmetric part by ( \ ), for
example$\;$ , $A_{(\mu \nu)}=(1/2)( A_{\mu \nu}+A_{\nu \mu})$ and
the antisymmetric part by the square bracket [\ ], $A_{[\mu
\nu]}=(1/2)( A_{\mu \nu}-A_{\nu \mu})$ .} \ba \A  \A {e^{(0)}}_0 =
1-{(2M \rho-Q^2+L^2)\rho^2 \over 2\rho_1}\; , \nonu
 \A \A {e^{(0)}}_\alpha = \left\{   -(n_\alpha-\displaystyle {a \over \rho}
 \epsilon_{\alpha  j 3 }\;  n^j)
 -\displaystyle {a^2 \over \rho^2}  \displaystyle {z \over \rho} {\delta_\alpha}^3
   \right\}{(2M \rho-Q^2+L^2)\rho^4  \over 2\rho_1(\rho^2+a^2)}=-{e^{(l)}}_0 \; ,    \nonu
\A  \A  {e^{(l)}}_\beta  ={\delta^l}_\beta+ \Biggl \{x^l x_\beta
-2\displaystyle {a \over \rho} \epsilon_{k 3 (\beta} x^{l)} x^k +
\displaystyle {a^2 \over \rho^2} \Biggl[ {\epsilon_{k}}^{l 3}
\epsilon_{m \beta 3} x^k x^m+ z\Biggl(\{\rho x^{l}-\displaystyle{a
\over \rho} {\epsilon_{k 3}}^{l} \; x^k \}{\delta_{\beta}}^3 \nonu
\A \A +\{\rho x_\beta-\displaystyle{a \over \rho} {\epsilon_{k 3
\beta}} \; x^k \} \delta^{l3} \Biggr)\Biggr]+\displaystyle{a^4
\over \rho^4}z^2 \delta_{\beta 3} \delta^{3 l} \Biggr \} {(2M
\rho-Q^2+L^2) \rho^4 \over 2\rho_1 (\rho^2+a^2)^2}\; , \ea with
${\delta^l}_\beta$ being the Kronecker delta \cite{KT} and
\[ \rho_1=\rho^4+a^2z^2\; .
\] Solution (22) with Eq. (20) satisfy the field equations
(7)$\sim$(9) and the associated metric has the following form \be
ds^2= dt^2-dx^2-dy^2-dz^2-\displaystyle{(2Mr_1-a^2-Q^2+L^2){r_1}^2
\over {r_1}^4+a^2z^2} \left\{dt+\displaystyle{zdz \over
r_1}+\displaystyle{r_1(xdx+ydy)+a(xdy-ydx) \over {r_1}^2+a^2}
\right\}^2,\ee with $r_1$ is the radial parameter related to the
Cartesian  radius $r$ by \[ {r_1}^4-(r^2-a^2){r_1}^2-a^2z^2=0.\]
\newsection{The energy and angular momentum associated with the general axi-symmetric solution}

 The superpotential is given by \cite{Mo1, MWHL}
  \be {{\cal U}_\mu}^{\nu \lambda} ={(-g)^{1/2} \over
2 \kappa} {P_{\chi \rho \sigma}}^{\tau \nu \lambda}
\left[\Phi^\rho g^{\sigma \chi} g_{\mu \tau}
 -\lambda g_{\tau \mu} \gamma^{\chi \rho \sigma}
-(1-2 \lambda) g_{\tau \mu} \gamma^{\sigma \rho \chi}\right]\; ,
\ee where ${P_{\chi \rho \sigma}}^{\tau \nu \lambda}$ is \be
{P_{\chi \rho \sigma}}^{\tau \nu \lambda} \stackrel{\rm def.}{=}
{{\delta}_\chi}^\tau {g_{\rho \sigma}}^{\nu \lambda}+
{{\delta}_\rho}^\tau {g_{\sigma \chi}}^{\nu \lambda}-
{{\delta}_\sigma}^\tau {g_{\chi \rho}}^{\nu \lambda} \ee with
${g_{\rho \sigma}}^{\nu \lambda}$ being a tensor defined by \be
{g_{\rho \sigma}}^{\nu \lambda} \stackrel{\rm def.}{=}
{\delta_\rho}^\nu {\delta_\sigma}^\lambda- {\delta_\sigma}^\nu
{\delta_\rho}^\lambda. \ee The energy-momentum density is defined
by \cite{Mo1} \be {\tau_\mu}^\nu={{{\cal U}_\mu}^{\nu
\lambda}}_{\; , \ \lambda}.\ee  The energy $E$  contained in a
sphere with radius $R$ is expressed by the volume  integral
   \cite{M58} \be
E(R)=\int_{r=R} \int \int   {\tau_0}^0 d^3x. \ee For convenience
of the calculations, we work for small values of the rotation
parameter $a$, so we will neglect the terms beyond its fourth
order. With this approximation, we have the covariant components
of the parallel vector fields (22) as
\\
\\
\ba {e^{(0)}}_0 \A=\A 1-\displaystyle{1 \over
r}\left(M-\displaystyle{Q^2-L^2 \over 2r}\right)-\displaystyle{a^2
\over 2r^3}\left[\left(M-\displaystyle{Q^2-L^2 \over
r}\right)-\left(3M-\displaystyle{2(Q^2-L^2) \over
r}\right)\displaystyle{z^2 \over r^2}\right]\nonu
\A \A -\displaystyle{a^4 \over
8r^5}\left[\left(3M-\displaystyle{4(Q^2-L^2) \over r}
\right)-\left(30M-\displaystyle{24(Q^2-L^2) \over r}
\right)\displaystyle{z^2 \over
r^2}+\left(35M-\displaystyle{24(Q^2-L^2) \over r}
\right)\displaystyle{z^4 \over r^4} \right],\nonu
 {e^{(0)}}_\alpha \A=\A -\left(M-\displaystyle{Q^2-L^2 \over 2r}\right)\displaystyle{n_\alpha \over
r}+\displaystyle{a \over r^2}\left(M-\displaystyle{Q^2-L^2 \over
2r}\right)\epsilon_{\alpha \beta 3} n^\beta + \displaystyle{a^2
\over r^3}\Biggl[\Biggl\{2\left(M-\displaystyle{Q^2-L^2 \over
2r}\right)\displaystyle{z^2 \over r^2}\nonu
\A \A+\displaystyle{Q^2-L^2 \over 4r^2}(x^2+y^2)\Biggr\}n_\alpha
-\left(M-\displaystyle{Q^2-L^2 \over 2r} \right)\displaystyle{z
\over r}{\delta_\alpha}^3\Biggr]+\displaystyle{a^3 \over
2r^4}\Biggl[\left(M-\displaystyle{Q^2-L^2 \over r}\right)\nonu
\A \A-\left(5M-\displaystyle{3(Q^2-L^2) \over
r}\right)\displaystyle{z^2 \over r^2}\Biggr]\epsilon_{\alpha \beta
3} n^\beta+\displaystyle{a^4 \over
8r^5}\Biggl[3\Biggl\{\displaystyle{Q^2-L^2 \over
2r}+\left(8M-\displaystyle{7(Q^2-L^2) \over r}\right)
\displaystyle{z^2 \over r^2} -\Biggl(16M \nonu
\A \A -\displaystyle{21(Q^2-L^2) \over 2r}\Biggr)\displaystyle{z^4
\over
r^4}\Biggr\}n_\alpha-2\Biggl\{2\left(2M-\displaystyle{3(Q^2-L^2)
\over 2r}\right)-\left(12M-\displaystyle{7(Q^2-L^2) \over
r}\right)\displaystyle{z^2 \over r^2}\Biggr\}\displaystyle{z \over
r} {\delta_\alpha}^3 \Biggr]=-\delta_{l \alpha}
{b^{(\alpha)}}_0,\nonu
 {e^{(l)}}_\alpha
\A=\A{\delta^l}_\alpha+\left(\displaystyle{n^l n_\alpha \over
r}-\displaystyle{2a \over r^2}{\epsilon^{(l}}_{\beta
3}n_{\alpha)}n^\beta \right)\left(M-\displaystyle{Q^2-L^2 \over
2r}\right)-\displaystyle{a^2 \over r^3}\Biggl[\displaystyle{1
\over 2}\left\{M+\left(5M-\displaystyle{3(Q^2-L^2) \over
r}\right)\displaystyle{z^2 \over r^2}\right\}n^l n_\alpha \nonu
\A \A -\left(M-\displaystyle{Q^2-L^2 \over
2r}\right){\epsilon^l}_{\beta 3} {\epsilon_{\alpha \gamma}}^
3n^\beta n^\gamma-\left(M-\displaystyle{Q^2-L^2 \over
2r}\right)\displaystyle{2z \over r}n^{(l} {\delta_{\alpha)}}^3
\Biggr] +\displaystyle{a^3 \over
r^4}\Biggl[\Biggl\{6\left(M-\displaystyle{Q^2-L^2 \over
2r}\right)\displaystyle{z^2 \over r^2}\nonu
\A \A+\displaystyle{Q^2-L^2 \over
2r^3}\left(x^2+y^2\right)\Biggr\}n^\beta{\epsilon^{(l}}_{\beta
3}n_{\alpha)}-\left(M-\displaystyle{Q^2-L^2 \over
2r}\right)\displaystyle{2z \over r}n^\beta {\epsilon^{(l}}_{\beta
3}{\delta_{\alpha)}}^3\Biggr]-\displaystyle{a^4 \over 8r^5}
\Biggl[\Biggl\{\left(5M-\displaystyle{4(Q^2-L^2) \over r}
\right)\nonu
\A \A-\Biggl(14M\displaystyle{z^2 \over r^2}
+\left(63M-\displaystyle{40(Q^2-L^2) \over r} \right)
\displaystyle{z^4 \over r^4} \Biggr) \Biggr\}n^l
n_\alpha-8\Biggl\{\left(M-\displaystyle{Q^2-L^2 \over
r}\right)\nonu
\A \A -\left(7M-\displaystyle{4(Q^2-L^2) \over r} \right)
\displaystyle{z^2 \over r^2}\Biggr\} \displaystyle{2z \over
r}n^{(l}{\delta_{\alpha)}}^3-4\Biggl\{\left(M-\displaystyle{Q^2-L^2
\over r}\right)-\left(8M-\displaystyle{5(Q^2-L^2) \over r} \right)
\displaystyle{z^2 \over r^2}\nonu
\A \A +\left(7M-\displaystyle{4(Q^2-L^2) \over r} \right)
\displaystyle{z^4 \over r^4}\Biggr\}{\delta^l}_\alpha
 +4\Biggl\{\left(M-\displaystyle{Q^2-L^2 \over
r}\right)-\left(9M-\displaystyle{5(Q^2-L^2) \over r} \right)
\displaystyle{z^2 \over r^2}\Biggr\} \delta^{3l} \delta_{3 \alpha}
\Biggr]\; ,
 \ea
and the contravariant components of the parallel vector fields
(22) have the form

 \ba {e_{(0)}}^0
\A=\A 1+\displaystyle{1 \over r}\left(M-\displaystyle{Q^2-L^2
\over 2r}\right)+\displaystyle{a^2 \over
2r^3}\left[\left(M-\displaystyle{Q^2-L^2 \over
r}\right)-\left(3M-\displaystyle{2(Q^2-L^2) \over
r}\right)\displaystyle{z^2 \over r^2}\right]\nonu
\A \A +\displaystyle{a^4 \over
8r^5}\left[\left(3M-\displaystyle{4(Q^2-L^2) \over r}
\right)-\left(30M-\displaystyle{24(Q^2-L^2) \over r}
\right)\displaystyle{z^2 \over
r^2}+\left(35M-\displaystyle{24(Q^2-L^2) \over r}
\right)\displaystyle{z^4 \over r^4} \right],\nonu
{e_{(0)}}^\alpha \A= \A \delta^{\alpha l}{b^{(0)}}_\alpha=
-{b_{(\alpha)}}^0 , \nonu
 {e_{(l)}}^\alpha
\A = \A{\delta_l}^\alpha-\left(\displaystyle{n_l n^\alpha \over
r}-\displaystyle{2a \over r^2}\epsilon_{\beta 3
(l}n^{\alpha)}n^\beta \right)\left(M-\displaystyle{Q^2-L^2 \over
2r}\right)+\displaystyle{a^2 \over r^3}\Biggl[\displaystyle{1
\over 2}\left\{M+\left(5M-\displaystyle{3(Q^2-L^2) \over
r}\right)\displaystyle{z^2 \over r^2}\right\}n_l n^\alpha \nonu
\A \A -\left(M-\displaystyle{Q^2-L^2 \over 2r}\right)\epsilon_{l
\beta 3} {\epsilon_{\gamma}}^{3 \alpha} n^\beta
n^\gamma-\left(M-\displaystyle{Q^2-L^2 \over
2r}\right)\displaystyle{2z \over r} n_{(l} \delta^{\alpha) 3}
\Biggr] -\displaystyle{a^3 \over
r^4}\Biggl[\Biggl\{6\left(M-\displaystyle{Q^2-L^2 \over
2r}\right)\displaystyle{z^2 \over r^2}\nonu
 \A \A +\displaystyle{Q^2-L^2
 \over 2r^3}\left(x^2+y^2\right)\Biggr\}n^\beta \epsilon_{\beta
3 (l}n^{\alpha)}-\left(M-\displaystyle{Q^2-L^2 \over
2r}\right)\displaystyle{2z \over r}n^\beta \epsilon_{\beta 3(l
}\delta^{\alpha) 3} \Biggr]+\displaystyle{a^4 \over 8r^5}
\Biggl[\Biggl\{\left(5M-\displaystyle{4(Q^2-L^2) \over r}
\right)\nonu
\A \A-\Biggl(14M\displaystyle{z^2 \over r^2}
+\left(63M-\displaystyle{40(Q^2-L^2) \over r} \right)
\displaystyle{z^4 \over r^4} \Biggr) \Biggr\}n_l
n^\alpha-8\Biggl\{\left(M-\displaystyle{Q^2-L^2 \over
r}\right)-\Biggl(7M-\displaystyle{4(Q^2-L^2) \over r} \Biggr)
\displaystyle{z^2 \over r^2}\Biggr\} \nonu
\A \A \displaystyle{2z \over r}n_{(l}\delta^{\alpha)
3}-4\Biggl\{\left(M-\displaystyle{Q^2-L^2 \over r}\right)
-\left(8M-\displaystyle{5(Q^2-L^2) \over r} \right)
\displaystyle{z^2 \over r^2}+\left(7M-\displaystyle{4(Q^2-L^2)
\over r} \right) \displaystyle{z^4 \over
r^4}\Biggr\}{\delta_l}^\alpha\nonu
\A \A +4\left\{\left(M-\displaystyle{Q^2-L^2 \over
r}\right)-\left(9M-\displaystyle{5(Q^2-L^2) \over r} \right)
\displaystyle{z^2 \over r^2}\right\}{\delta_{3l}}{\delta^{3
\alpha}} \Biggr]. \ea Also with this approximation the covariant
components of the metric tensor have the form
 \ba g_{0 0} \A=\A
1-\displaystyle{2 \over r}\left(M-\displaystyle{Q^2-L^2 \over
2r}\right)-\displaystyle{a^2 \over
r^3}\left[\left(M-\displaystyle{Q^2-L^2 \over
r}\right)-\left(3M-\displaystyle{2(Q^2-L^2) \over
r}\right)\displaystyle{z^2 \over r^2}\right]\nonu
\A \A -\displaystyle{a^4 \over
4r^5}\left[\left(3M-\displaystyle{4(Q^2-L^2) \over r}
\right)-\left(30M-\displaystyle{24(Q^2-L^2) \over r}
\right)\displaystyle{z^2 \over
r^2}+\left(35M-\displaystyle{24(Q^2-L^2) \over r}
\right)\displaystyle{z^4 \over r^4} \right],\nonu
g_{0 \alpha} \A=\A  -\left(M-\displaystyle{Q^2-L^2 \over
2r}\right){2n_\alpha \over r}+\displaystyle{2a \over
r^2}\left(M-\displaystyle{Q^2-L^2 \over 2r}\right)\epsilon_{\alpha
\beta 3} n^\beta+\displaystyle{2a^2 \over
r^3}\Biggl[\Biggl\{2\left(M-\displaystyle{Q^2-L^2 \over
2r}\right)\displaystyle{z^2 \over r^2}+\displaystyle{Q^2-L^2 \over
4r^2}\nonu
\A \A(x^2+y^2)\Biggr\}n_\alpha -\left(M-\displaystyle{Q^2-L^2
\over 2r} \right)\displaystyle{z \over
r}{\delta_\alpha}^3\Biggr]+\displaystyle{a^3 \over
r^4}\left[\left(M-\displaystyle{Q^2-L^2 \over
r}\right)-\left(5M-\displaystyle{3(Q^2-L^2) \over
r}\right)\displaystyle{z^2 \over r^2}\right]\epsilon_{\alpha \beta
3} n^\beta\nonu
\A \A+\displaystyle{a^4 \over
4r^5}\Biggl[3\Biggl\{\displaystyle{Q^2-L^2 \over
2r}+\left(8M-\displaystyle{7(Q^2-L^2) \over r}\right)
\displaystyle{z^2 \over r^2}  -\left(16M-\displaystyle{21(Q^2-L^2)
\over 2r}\right)\displaystyle{z^4 \over r^4}\Biggr\}n_\alpha\nonu
\A \A -2\left\{2\left(2M-\displaystyle{3(Q^2-L^2) \over
2r}\right)-\left(12M-\displaystyle{7(Q^2-L^2) \over
r}\right)\displaystyle{z^2 \over r^2}\right\}\displaystyle{z \over
r} {\delta_\alpha}^3 \Biggr],\nonu
g_{\alpha \beta} \A=\A -\delta_{\alpha
\beta}-2\left(\displaystyle{  n_\alpha n_\beta \over
r}-\displaystyle{2a \over r^2}\epsilon_{ \gamma 3 (\beta
}n_{\alpha)}n^\gamma \right)\left(M-\displaystyle{Q^2-L^2 \over
2r}\right)+\displaystyle{a^2 \over
r^3}\Biggl[\left\{M+\left(5M-\displaystyle{3(Q^2-L^2) \over
r}\right)\displaystyle{z^2 \over r^2}\right\}n_\beta n_\alpha
\nonu
\A \A +2\left(M-\displaystyle{Q^2-L^2 \over 2r}\right)\epsilon_{
\epsilon \beta 3} {\epsilon_{\alpha \gamma}}^3n^\epsilon
n^\gamma-4\left(M-\displaystyle{Q^2-L^2 \over
2r}\right)\displaystyle{z \over r}n_{(\beta
}{\delta_{\alpha)}}^3\Biggr] -\displaystyle{a^3 \over
r^4}\Biggl[\Biggl\{12\left(M-\displaystyle{Q^2-L^2 \over
2r}\right)\displaystyle{z^2 \over r^2}\nonu
\A \A+\displaystyle{Q^2-L^2 \over
r^3}\left(x^2+y^2\right)\Biggr\}n^\gamma
n_{(\alpha}-4\left(M-\displaystyle{Q^2-L^2 \over 2r}\right)
\displaystyle{z \over r}n^\gamma
{\delta_{(\alpha}}^3\Biggr]\epsilon_{\beta)  \gamma 3
}+\displaystyle{a^4 \over 4r^5}
\Biggl[\Biggl\{\left(5M-\displaystyle{4(Q^2-L^2) \over r}
\right)\nonu
\A \A -\Biggl(14M\displaystyle{z^2 \over
r^2}+\left(63M-\displaystyle{40(Q^2-L^2) \over r} \right)
\displaystyle{z^4 \over r^4} \Biggr) \Biggr\}n_\beta
n_\alpha-16\left\{\left(M-\displaystyle{Q^2-L^2 \over
r}\right)-\left(7M-\displaystyle{4(Q^2-L^2) \over r} \right)
\displaystyle{z^2 \over r^2}\right\} \nonu
\A \A \displaystyle{z \over r}n_{(\beta}{\delta_{\alpha)}}^3
-4\Biggl\{\left(M-\displaystyle{Q^2-L^2 \over
r}\right)-\left(8M-\displaystyle{5(Q^2-L^2) \over r} \right)
\displaystyle{z^2 \over r^2}+\left(7M-\displaystyle{4(Q^2-L^2)
\over r} \right) \displaystyle{z^4 \over r^4}\Biggr\}\delta_{\beta
\alpha} \nonu
\A \A +4\left\{\left(M-\displaystyle{Q^2-L^2 \over
r}\right)-\left(9M-\displaystyle{5(Q^2-L^2) \over r} \right)
\displaystyle{z^2 \over r^2}\right\}\delta_{\beta 3}\delta_{3
\alpha} \Biggr],
 \ea
and the asymptotic form of the electromagnetic potential $A_\mu$
has the form  \be A_0= -\frac{q}{4 \ r \pi}+O
\left(\frac{1}{r^3}\right), \qquad
A_\alpha=-\frac{q}{4\pi}\frac{x^\alpha}{r^2}+O_\alpha\left(\frac{1}{r^3}\right).\ee
The determinant of the metric tensor is \be g=-1.\ee Now we
evaluate all the required components of the contorsion and the
basic vector using ((2) and (4)) neglecting terms beyond the
fourth powers of ${\it a}$.

 Substituting Eq. (31) and all components of the contorsion and basic vector in (24)
we get \ba {{\cal U}_0}^{0 \alpha} \A =\A {2n^\alpha \over \kappa
r^2} \Biggl[(M-\displaystyle{Q^2-L^2 \over 2r})+\displaystyle{a^2
\over r^2}\left\{M(1-\displaystyle{5z^2 \over
r^2})-\displaystyle{Q^2-L^2 \over r}(1-\displaystyle{3z^2 \over
r^2}) \right\}+\displaystyle{a^4 \over 8r^4}\Biggl(M\Biggl\{(9-126
\displaystyle{z^2 \over r^2}+ \displaystyle{189z^4 \over
r^4})\Biggr\} \nonu
\A \A -\displaystyle{Q^2-L^2 \over r}\Biggl\{
(12-\displaystyle{96z^2 \over r^2}+\displaystyle{120z^4 \over
r^4}) \Biggr\}\Biggr)\Biggr]+{4z {\delta^\alpha}_3 \over \kappa
r^3} \Biggl[ \displaystyle{a^2 \over r^2}(M-\displaystyle{Q^2-L^2
\over 2r}) +\displaystyle{a^4 \over 4r^4}\Biggl(M(9-
\displaystyle{21z^2 \over r^2})\nonu
\A \A-6\displaystyle{Q^2-L^2 \over r}(1- \displaystyle{2z^2 \over
r^2})\Biggr)\Biggr]-{M n_\beta \epsilon^{\alpha \beta 3} \over
\kappa r^2}\Biggl(\displaystyle{a \over r}+\displaystyle{a^3 \over
r^3}(1-3\displaystyle{z^2 \over r^2})\Biggr)\; , \ea

 and \ba {{\cal U}_\alpha}^{0 \beta} \A =\A {1 \over 2
\kappa r^2}\Biggl[\displaystyle{Q^2-L^2 \over r}
{\delta_\alpha}^\beta+2\left(2M-\displaystyle{3(Q^2-L^2) \over
2r}\right)n_\alpha n^\beta -\displaystyle{a \over
r}\Biggl\{\left(2M-\displaystyle{Q^2-L^2 \over
r}\right){\epsilon_{\alpha 3}}^{ \beta}+2\Biggl(3M\nonu
\A \A -\displaystyle{2(Q^2-L^2) \over r}\Biggr)  \epsilon_{\alpha
\gamma 3} n^\gamma n^\beta \Biggr \}+\displaystyle{a^2 \over
r^2}\Biggl\{2\left(M-\displaystyle{Q^2-L^2 \over
r}\right)\left({\delta_\alpha}^\beta-(1-3\displaystyle{Q^2-L^2
\over 2r}){\delta_\alpha}^3{\delta_3}^\beta\right)-\Biggl(24M
\nonu
\A \A -\displaystyle{35(Q^2-L^2) \over 2r}\Biggr)\displaystyle{z^2
\over r^2} n_\alpha n^\beta+\displaystyle{5(Q^2-L^2) \over
2r}\epsilon_{\alpha \gamma 3}{\epsilon_\delta}^{3 \beta} n^\gamma
n^\delta +2\left(8M-\displaystyle{15(Q^2-L^2) \over
2r}\right)\displaystyle{z \over r}
n^{(\beta}{\delta_{\alpha)}}^3\Biggr \} \nonu
\A \A-\displaystyle{a^3 \over
r^3}\Biggl\{\left(M-\displaystyle{Q^2-L^2 \over
r}\right){\epsilon_{\alpha 3}}^\beta
+\Biggl[\left(5M-\displaystyle{6(Q^2-L^2) \over
r}\right)-\left(35M-\displaystyle{24(Q^2-L^2) \over r}\right)
\displaystyle{z^2 \over r^2} \Biggr]\Biggl(\epsilon_{\alpha \gamma
3}n^\gamma n^\beta\nonu
\A \A+\left(10M-\displaystyle{6(Q^2-L^2) \over
r}\right)\displaystyle{z \over r} n^\rho \; \epsilon_{\alpha \rho}
\; {\delta^\beta}_3 \Biggr)-\left(5M-\displaystyle{3(Q^2-L^2)
\over r}\right)\displaystyle{z^2 \over r^2} {\epsilon_{\alpha
3}}^\beta \Biggr\}+\displaystyle{a^4 \over
r^4}\Biggl\{\Biggl[\left(2M-\displaystyle{9(Q^2-L^2) \over
4r}\right) \nonu
\A \A-\left(54M-\displaystyle{357(Q^2-L^2) \over
8r}\right)\displaystyle{z^2 \over
r^2}+\Biggl(168M-\displaystyle{126(Q^2-L^2) \over
r}\Biggr)\displaystyle{z^4 \over
r^4}-\left(120M-\displaystyle{693(Q^2-L^2) \over
8r}\right)\displaystyle{z^6 \over r^6}\Biggr]\nonu
\A \A\Biggl({\delta_\alpha}^\beta-(1-15\displaystyle{Q^2-L^2 \over
8r}){\delta_\alpha}^3{\delta_3}^\beta\Biggr)
+\Biggl[\displaystyle{21(Q^2-L^2) \over
8r}+\Biggl(48M-\displaystyle{189(Q^2-L^2) \over
4r}\Biggr)\displaystyle{z^2 \over r^2} -\Biggl(120M
 \nonu
\A \A-\displaystyle{693(Q^2-L^2) \over 8r}\Biggr)\displaystyle{z^4
\over r^4} \Biggr]\epsilon_{\alpha \gamma
3}{\epsilon_\delta}^{3\beta}n^\gamma
n^\delta+2\Biggl[\Biggl(12M-\displaystyle{105(Q^2-L^2) \over
8r}\Biggr)-\left(96M-\displaystyle{315(Q^2-L^2) \over
4r}\right)\displaystyle{z^2 \over r^2}
 \nonu
\A \A+\Biggl(120M-\displaystyle{693(Q^2-L^2) \over
8r}\Biggr)\displaystyle{z^4 \over r^4} \Biggr] \displaystyle{z
\over r}
n^{(\beta}{\delta_{\alpha)}}^3+\Biggl[\Biggl(12M-\displaystyle{105(Q^2-L^2)
\over 8r}\Biggr)\displaystyle{z^2 \over r^2}\nonu
\A \A +\left(48M-\displaystyle{315(Q^2-L^2) \over
4r}\right)\displaystyle{z^4 \over r^4}-\Biggl(120M
-\displaystyle{693(Q^2-L^2) \over 8r}\Biggr)\displaystyle{z^6
\over r^6}\Biggr] \delta_{\alpha 3} \delta^{\beta
3}\Biggr\}\Biggr].
 \ea

The components we are interested in of the energy-momentum density
are given by \ba {\tau_0}^0 \A=\A \displaystyle{Q^2-L^2 \over
\kappa r^4} \left[1- \displaystyle{a^2 \over  r^2}\left(2-
\displaystyle{6 \over r^2}(x^2+y^2)\right)+ \displaystyle{a^4
\over r^4}\left(3-\displaystyle{24 \over
r^2}(x^2+y^2)+\displaystyle{30 \over r^4}(x^2+y^2)^2\right)
\right]\; ,\nonu
{\tau_\alpha}^0 \A=\A -\displaystyle{2 a (Q^2-L^2)
\epsilon_{\alpha \beta 3} n^\beta \over \kappa
r^5}\left(1+\displaystyle{3a^2 \over r^2}(r^2-2z^2) \right). \ea

Further substituting (36) in (28) and then transforming it into
spherical coordinate we obtain \be E=\displaystyle{(Q^2-L^2) \over
\kappa r^6}\int_{r=R} \int \int \left(r^4+4a^2r^2-6a^2r^2 \cos^2
\theta+9a^4-36a^4 \cos^2 \theta+30a^4 \cos^4 \theta\right) dr \sin
\theta d\theta d\phi.\ee

Performing the above integration we get the energy associated with
the exterior general axisymmetric black hole given by Eq. (22) in
the form \be {E(R)_{total}}^{exterior}=\displaystyle{(Q^2-L^2)
\over R} \left(\displaystyle{1 \over 2}+\displaystyle{a^2 \over
3R^2}+\displaystyle{3a^4 \over 10R^4}
\right)+O\left(\frac{1}{R^6}\right).\ee Eq. (38) shows that the
total energy associated with the spacetime given by Eq. (22) is
shared by its exterior as well as interior.

 Now we turn our attention to the angular momentum of the
 solution given by Eq. (22). From (36) we can get the components of the
 momentum density in the form
 \be P_\alpha= -\displaystyle{2 a (Q^2-L^2)\epsilon_{\alpha \beta 3} n^\beta \over
\kappa r^5}\left(1+\displaystyle{3a^2 \over r^2}(r^2-2z^2)
\right)\; ,\ee from which we can show that there is no
momentum-density associated with Kerr black hole i.e,. when
$Q=L=0$. The components of the angular momentum of a
general-relativistic system is given by \cite{VM} \be J_\alpha=
\int \int \int (x_\beta P_\gamma-x_\gamma P_\beta) d^3 x\; ,\ee
where $\alpha$, $\beta$, $\gamma$ take cyclic values 1,2,3. Using
(39) in (40) and transforming the expressions into spherical
coordinates
 \be J_\alpha=\int \int \int \displaystyle{2 a (Q^2-L^2)  \over
\kappa r^4} \left(r^2+3a^2(1-2\cos^2\theta \right)\sin^3\theta dr
d\theta d\phi\; ,\ee performing the above integration between two
spheres of radii $R_1$ and $R_2$  we get \be J_\alpha=-2a(Q^2-L^2)
\epsilon_{12\alpha} \left[\displaystyle{1 \over
3r}+\displaystyle{a^2 \over 5r^3} \right]_{R_1}^{R_2}.\ee Eq. (42)
shows that the rotation of the charged object  is responsible for
the angular momentum distribution  due to the electromagnetic
field  and the NUT parameter.
\newsection{Main results and discussion}

In this paper we have studied the coupled equations of the
gravitational and electromagnetic fields  in the tetrad theory of
gravitation. Applying the most general  tetrad (15) to the field
equations (7)$\sim$(9) we have obtained an exact  solution given
by Eq. (22). Eq. (22) is a general axi-symmetric solution from
which we can generate  the other known solutions like
Schwarzschild, Reissner-Nordstr$\ddot{o}$m, Kerr and Kerr-Newman
spacetimes. The metric tensor associated with Eq. (22) is the
Kerr-NUT spacetime.

It was shown by M\o ller \cite{Mo26} that the tetrad description
of the gravitational field allows a more satisfactory treatment of
the energy-momentum complex than does GR. Therefore, we have used
the superpotential (24), {\it formulated within the framework of
teleparallel spacetime} to calculate the energy-momentum density
 given by Eq. (27). We have used this energy-momentum density to evaluate the
total exterior energy of the gravitating charged rotating body
given by Eq. (22). Because the definition of energy of Eq. (28)
requires its evaluation in Cartesian coordinate, the calculations
without any approximation is obviously very tedious. Moreover, the
intrinsic rotation parameter $a$ is quantitatively very small for
most physical situations. Therefore, for our convenience we keep
terms containing powers of $a$ up to the fourth order. With this
approximation we have calculated the components of a covariant and
contravariant tetrad fields (29) and (30), a covariant components
of metric tensor (31). Calculating all the necessary components of
the contorsion and basic vector ((2) and (4))  and using (31) in
(28) we have obtained the expression of the exterior energy of the
general black hole given by Eq. (22) till the fourth order. When
the rotation, the charge as well as the NUT parameters are
considered we get an additional terms \be \displaystyle{Q^2-L^2
\over R} \left(\displaystyle{1 \over 2}+\displaystyle{a^2 \over
3R^2}+\displaystyle{3a^4 \over 10R^4} \right),\ee which is the
energy of the exterior magnetic field due to the rotation of the
charged object. The asymptotic value of the total gravitational
mass of a Kerr-NUT spacetime is the ADM \cite{ADM} therefore, the
energy associated with a Kerr-NUT contained in a sphere of radius
$R$ is \be E(R)=M-\displaystyle{Q^2-L^2 \over R}
\left(\displaystyle{1 \over 2}+\displaystyle{a^2 \over
3R^2}+\displaystyle{3a^4 \over 10R^4} \right). \ee  Switching off
the rotation and NUT parameters we found that the energy given by
Eq. (44)  will be the same as  of Reissner-Nordstr$\ddot{o}$m
metric \cite{Vc,Tp}. The energy of Eq. (44) is confined to its
interior only when we set the charge and the NUT parameters to be
zero. This result is quit in conformity with those of Virbhadra
\cite{Vs,Vs1} and Cooperstock et al. \cite{CR} and Ahmed
\cite{AH}.

Using the value of energy-momentum density given by Eq. (36) we
have calculated the momentum density.  Also we have calculated the
angular momentum distribution due to the electromagnetic field
present in the Kerr-NUT field using Eq. (39) in Eq. (40). It is
clear from Eq. (42) that the momentum density depends mainly on
the rotation of the charged object and has only one component.  As
is clear from Eq. (42) that the angular momentum depends on the
even powers of the charge and NUT parameters and odd powers of the
rotation parameter which means that the direction of the angular
momentum vector depends on the direction of the rotation
parameter.

\vspace{2cm} \centerline{\large {\bf Acknowledgment}} The author
would like to  thank the Referee for careful reading and for
checking the mathematics of the manuscript as well as for putting
the paper in a more readable form .

\end{document}